\documentclass[12pt]{article}

\usepackage{epsf,epsfig,psfig}

\def\spose#1{\hbox to 0pt{#1\hss}}
\def\ltapprox{\mathrel{\spose{\lower 3pt\hbox{$\mathchar"218$}}
 \raise 2.0pt\hbox{$\mathchar"13C$}}}
\def\gtapprox{\mathrel{\spose{\lower 3pt\hbox{$\mathchar"218$}}
 \raise 2.0pt\hbox{$\mathchar"13E$}}}
\def\case#1#2{{\textstyle\frac{#1}{#2}}}

\begin{document}
\begin{titlepage}
\begin{flushright}
CERN-PH-TH/2004-029
\end{flushright}
\begin{center}
{\Large\bf 
Critical slowing down of topological modes
}
\end{center}
\vskip 1.3cm
\centerline{
Luigi Del Debbio,$^a$ 
Gian Mario Manca,$^b$ 
Ettore Vicari$\,^b$}

\vskip 0.4cm
\centerline{\sl  $^a$ CERN, Department of Physics, TH Division, CH-1211 Geneva 23}
\centerline{\sl  $^b$ Dipartimento di Fisica dell'Universit\`a di Pisa,
                       Pisa, Italy}

\vskip 1.cm

\begin{abstract}
 We investigate the critical slowing down of the topological modes
using local updating algorithms in lattice 2-$d$ $\mathrm{CP}^{N-1}$
models.  We show that the topological modes experience a critical
slowing down that is much more severe than the one of the
quasi-Gaussian modes relevant to the magnetic susceptibility, which is
characterized by $\tau_{\rm mag} \sim \xi^z$ with $z\approx 2$.  We
argue that this may be a general feature of Monte Carlo simulations of
lattice theories with non-trivial topological properties, such as QCD,
as also suggested by recent Monte Carlo simulations of 4-$d$ SU($N$)
lattice gauge theories.
\end{abstract}

\end{titlepage}

Monte Carlo simulations of critical phenomena in statistical mechanics
and of quantum field theories, such as QCD, in the continuum limit are
hampered by the problem of critical slowing down (CSD)
\cite{Sokal-92}.  The autocorrelation time $\tau$, which is related to
the number of iterations needed to generate a new independent
configuration, grows with increasing length scale $\xi$.  In
simulations of lattice QCD where the upgrading methods are essentially
local, it has been observed that the autocorrelation times of
topological modes are typically much larger than those of other
observables not related to topology, such as Wilson loops and their
correlators, see for instance
Refs. \cite{ABDDV-96}--\cite{Leinweber:2003sj}.  Recent Monte Carlo simulations
\cite{DPV-02,DPRV-02} of the 4-$d$ SU($N$) lattice gauge theories (for
$N=3,4,6$) provided evidence of a severe CSD for the topological
modes, using a rather standard local overrelaxed upgrading algorithm
(constructed taking a mixture of overrelaxed microcanonical and
heat-bath updatings).  Indeed, the autocorrelation time $\tau_Q$ of
the topological charge grows very rapidly with the length scale
$\xi\equiv \sigma^{-1/2}$, where $\sigma$ is the string tension,
showing an apparent exponential behavior $\tau_Q\sim \exp (c\xi)$ in
the range of values of $\xi$ where data are available.  This peculiar
effect was not observed in plaquette--plaquette or Polyakov line
correlations, suggesting an approximate decoupling between topological
modes and non-topological ones, such as those determining the
confining properties.  The issue of the CSD of topological modes is
particularly important for lattice QCD, because it may pose a serious
limitation for numerical studies of physical issues related to
topological properties, such as the mass and the matrix elements of
the $\eta'$ meson, and in general the physics related to the broken
U(1)$_A$ symmetry.

The above-mentioned results suggest that the dynamics of the
topological modes in Monte Carlo simulations is rather different from
that of quasi-Gaussian modes.  CSD of quasi-Gaussian modes for
traditional local algorithms, such as standard Metropolis or heat
bath, is related to an approximate random-walk spread of information
around the lattice.  Thus, the corresponding autocorrelation time
$\tau$ is expected to behave as $\tau\sim\xi^2$ (an independent
configuration is obtained when the information travels a distance of
the order of the correlation length $\xi$, and the information is
transmitted from a given site/link to the nearest neighbors).  This
guess is correct for Gaussian (free-field) models; in general it is
expected that $\tau\sim \xi^z$, where $z$ is a dynamical critical
exponent, and $z\approx 2$ for quasi-Gaussian modes.~\footnote{
Optimized overrelaxation procedures may achieve a reduction of $z$,
although the condition $z\ge 1$ holds for local algorithms
\cite{Adler-88}.} On the other hand, in the presence of relevant
topological modes, the random-walk picture may fail, and therefore we
may have qualitatively different types of CSD. These modes could give
rise to sizeable free-energy barriers separating different regions of
the configuration space.  The evolution in this space would then
present a long-time relaxation due to transitions between different
topological charge sectors, and the corresponding autocorrelation time
should behave as $\tau_{\rm top}\sim \exp F_b$, where $F_b$ is the
typical free-energy barrier between different topological sectors.
However, for this picture to become more quantitative, one should
understand how the typical free-energy barriers scale with the
correlation length. For example, we may still have a power-law
behavior if $F_b \sim \ln \xi$, or an exponential behavior if $F_b\sim
\xi^\theta$.  It is worth mentioning that in physical systems, such as
random-field Ising systems \cite{Natterman98} and glass models
\cite{Bouchaud98}, the presence of significant free-energy barriers in
the configuration space causes a very slow dynamics, and an effective
separation of short-time relaxation within the free-energy basins from
long-time relaxation related to the transitions between basins.  In
the case of random-field Ising systems the free-energy barrier picture
supplemented with scaling arguments leads to the prediction that
$\tau\sim \exp (c \xi^\theta)$, where $\theta$ is a universal critical
exponent \cite{Fisher-86}.

Motivated by the recent results of Ref.~\cite{DPV-02}, suggesting an
exponential CSD for the topological modes in 4-$d$ SU($N$) lattice
gauge theories, we decided to investigate this issue in 2-d
$\mathrm{CP}^{N-1}$ models~\cite{DDL-79,Witten-79}, where we can study
in detail the dependence of the autocorrelation time on the length
scale $\xi$ as $N$ is varied. Since the 2-d $\mathrm{CP}^{N-1}$ models
possess interesting properties expected to hold in QCD, such as
asymptotic freedom and a non-trivial topological structure, they have
often been used as a theoretical laboratory.  In particular, their
lattice formulation has been considered to check and develop methods
to investigate topological properties in asymptotically free models,
exploiting also large-$N$ analytic calculations.  See, e.g.,
Refs.~\cite{CR-r}--\cite{ADGL-03}. The CSD of the topological modes
in lattice $\mathrm{CP}^{N-1}$ models, and in particular the behavior
of the autocorrelation time of the topological susceptibility, has
already been discussed in Refs.~\cite{CRV-92,V-93}, where the
hypothesis of a strong CSD was put forward on the basis of a few rough
estimates of $\tau_{\rm top}$ for the $\mathrm{CP}^9$ model, and the
fact that for large $N$, $N=O(100)$ say, it was not possible to
correctly sample the topological sectors.  In this paper we present
high-statistics Monte Carlo simulations using local updating
algorithms, such as Metropolis and overrelaxed algorithms, obtaining
rather accurate estimates of the topological susceptibility and its
integrated autocorrelation time.  The results provide a definite
evidence that, under local updating algorithms, the CSD experienced by
the topological modes turns out to be much more severe than the CSD of
the magnetic susceptibility, whose autocorrelation time shows a
power-law behavior $\tau_{\rm mag}\sim \xi^z$ with $z\approx 2$.

Two-dimensional $\mathrm{CP}^{N-1}$ models are defined by the action
\begin{equation}
S= {N\over g} \int d^2x\,\overline{D_\mu z} D_\mu z,
\end{equation}
where $z$ is an $N$-component complex scalar field subject to the
constraint $\bar{z}z=1$, and the covariant derivative $D_\mu
=\partial_\mu +iA_\mu$ is defined in terms of the composite field
$A_\mu=i\bar{z}\partial_\mu z$.  Like QCD, they are asymptotically
free and present non-trivial topological structures (instantons,
anomalies, $\theta$ vacua).  The large-$N$ expansion is performed by
keeping the coupling $g$ fixed \cite{DDL-79,Witten-79}.  A topological
charge density operator $q(x)$ can be defined as
\begin{equation}
q(x)={1\over2\pi}\epsilon_{\mu\nu} \partial_\mu A_\nu,
\label{contch}
\end{equation}
with the related topological susceptibility
\begin{equation}
\chi_t = \int d^2x\langle q(x)q(0) \rangle.
\end{equation}
We consider the lattice formulation \cite{RS-81,DHMNP-81,BL-81,CRV-92}
\begin{eqnarray}
S_L &=& - N \beta \Bigl[ {4\over 3}
\sum_{n,\mu}\left( 
   \bar z_{n+\mu}z_n\lambda_{n,\mu} +
   \bar z_nz_{n+\mu}\bar\lambda_{n,\mu} - 2\right) 
\label{sla}
\\ &&
-{1\over 12} \sum_{n,\mu}\left( 
   \bar z_{n+2 \mu}z_n\lambda_{n,\mu}\lambda_{n+\mu,\mu} +
   \bar z_nz_{n+2\mu}\bar\lambda_{n,\mu} \bar\lambda_{n+\mu,\mu} -
 2\right) \Bigr],
\nonumber
\end{eqnarray}
where, beside the complex $N$-component vector $z$ satisfying $\bar{z}
z=1$, the complex variable $\lambda_{n,\mu}$ has been introduced,
which satisfies $\bar{\lambda}_{n,\mu}\lambda_{n,\mu}=1$;  $S_L$ is a
tree-order Symanzik-improved lattice action \cite{Symanzik,CRV-92}.
The correlation function is defined as
\begin{equation}
G(x) = \langle {\rm Tr} P(x) P(0) \rangle_{\rm conn},
\end{equation}
where $P = \bar{z} \otimes z $.  One can define the magnetic
susceptibility $\chi_m$ and the second-moment correlation length $\xi$
from its small-momentum behavior:
\begin{equation}
\chi_m = \widetilde{G}(0),
\quad
\xi^2 = {1\over 4 \sin^2(q_{\rm m}/2)} 
{ \widetilde{G}(0) - \widetilde{G}(q_{\rm m})\over 
\widetilde{G}(q_{\rm m}) },
\end{equation}
where $q_{\rm m}=(2\pi/L,0)$ is the minimum non-zero momentum on a
lattice of size $L$ with periodic boundary conditions (see
Ref.~\cite{CP-98} for a discussion of this estimator of the
second-moment correlation length).  We consider the geometrical
definition of lattice topological charge proposed in
Ref.~\cite{BL-81}, which meets the demands that the topological charge
on the lattice have the classical correct continuum limit and be an
integer for every lattice configuration in a finite volume with
periodic boundary conditions. As a result both the topological charge
and its susceptibility do not require lattice renormalizations. It is
given by \cite{BL-81}
\begin{equation}
Q = \sum_n 
{1\over 2 \pi} 
{\rm Im}\left[ \ln{\rm Tr}(P_{n+\mu+\nu}
P_{n+\mu}P_n)+ \ln{\rm Tr}
(P_{n+\nu}P_{n+\mu+\nu}P_n)\right],
\qquad \mu \neq \nu, 
\label{qgeomP}
\end{equation}
where the imaginary part of the logarithm is to be taken in $(-\pi,
\pi)$.  As shown in Refs.~\cite{CRV-92,RRV-97}, this definition is
effective for sufficiently large values of $N$, where unphysical
dislocations \cite{Luscher-82} should not affect the continuum limit
of its matrix elements.  The corresponding topological susceptibility
is obtained by
\begin{equation}
\chi_t = {1\over V} \langle Q^2 \rangle,
\end{equation}
where $V$ is the volume of the lattice.

The autocorrelation function $C_O(t)$ ($t$ is the discrete Monte Carlo
time, where a time unit is given by a sweep, i.e.  an update of all
lattice variables) of a given quantity $O$ is defined as
\begin{equation}
C_O(t) = \left\langle \left( O(t) - \langle O \rangle \right)
\left( O(0) - \langle O \rangle \right) \right\rangle,
\end{equation}
where the averages are taken at equilibrium. The integrated
autocorrelation time $\tau_O$ associated with $O$ is given by
\begin{equation}
\tau_O = {1\over 2} \sum_{t=-\infty}^{t=+\infty} {C_O(t)\over C_O(0)}.
\end{equation}
Estimates of $\tau_O$ can be obtained by the binning method (see
e.g. Ref.~\cite{Wolff-04} for a discussion of this method and its
systematic errors), using the estimator
\begin{equation}
\tau_O = {E^2\over 2 E_0^2},
\label{eqbl}
\end{equation}
where $E_0$ is the naive error calculated without taking into account
the autocorrelations, and $E$ is the error found after binning,
i.e. when the error estimate becomes stable with respect to increasing
the block size $n_b$.  The statistical error $\Delta\tau_O$ is just
given by $\Delta \tau_O/\tau_O = \sqrt{2/n_{b}}$, where $n_{b}$ is the
number of blocks corresponding to the estimate of $E$.  As discussed
in Ref.~\cite{Wolff-04} this procedure leads to a systematic error of
$O(\tau_O/b)$, where $b$ is the size of the blocks.  In our cases the
ratio $\tau_O/b$ will always be much smaller than the statistical
error, so we will neglect it.  Equation~(\ref{eqbl}) can be easily
extended to the case where the quantity $O$ is measured every $n_m$
sweeps, i.e. $\tau_O= n_m E^2/(2 E_0^2)$, which is of course
meaningful only if $n_m\ll \tau_O$.

We performed Monte Carlo simulations for $N=10,15,21$, for which the
geometrical definition (\ref{qgeomP}) should be effective to describe
the topological modes relevant to the continuum limit.  We measured
the magnetic susceptibility $\chi_m$, the correlation length $\xi$,
the topological susceptibility $\chi_t$, and the integrated
autocorrelation times of $\chi_m$ and $\chi_t$, respectively
$\tau_{\rm mag}$ and $\tau_{\rm top}$.  We considered two types of
updating methods: a standard Metropolis and a mixed method containing
overrelaxation procedures.  A summary of our runs is reported in
Table~\ref{tableres}.  Finite-size effects in lattice
$\mathrm{CP}^{N-1}$ models are rather large and peculiar, especially
at large $N$ \cite{RV-93}.  We performed our simulations for lattice
size $L$ sufficiently large to guarantee that finite-size effects were
at most $O(10^{-3})$ for $\xi$ and smaller than 1\% for $\chi_t$,
i.e. $L/\xi\gtapprox 10$ for $N=10$, $L/\xi\gtapprox 13$ for $N=15$,
and $L/\xi\gtapprox 15$ for $N=21$.  Each run consisted typically of a
few million sweeps for the smallest values of $\beta$, increasing up
to approximately 50 million for the largest $\beta$'s.

\begin{table*}
\caption{ Summary of the Monte Carlo data. The Metropolis and the
mixed overrelaxed upgrading methods are indicated by mt and ov
respectively.  }
\label{tableres}
\tiny
\begin{center}
\begin{tabular}{ccccllc}
\hline\hline
\multicolumn{1}{c}{$N$}&
\multicolumn{1}{c}{$\beta$}&
\multicolumn{1}{c}{$L$}&
\multicolumn{1}{c}{upgrade}&
\multicolumn{1}{c}{$\xi$}&
\multicolumn{1}{c}{$\chi_t \xi^2$}&
\multicolumn{1}{c}{$\tau_{\rm top}$}\\
\hline\hline 
10 & 0.59 & 20 & mt & 1.854(8) & 0.02223(24) & 17.7(6) \\

   & 0.61 & 24 & mt & 2.113(6) & 0.02181(19) & 31.1(1.0) \\
   & 0.61 & 24 & ov & 2.118(3) & 0.02204(11) & 3.44(10) \\

   & 0.63 & 30 & mt & 2.410(5) & 0.02145(10) & 53.3(1.9) \\
   & 0.63 & 30 & ov & 2.409(3) & 0.02117(14) & 5.6(3)\\

   & 0.65 & 32 & mt & 2.748(5) & 0.02048(10) & 83(3) \\
   & 0.65 & 32 & ov & 2.7471(12) & 0.02064(6)& 7.8(2)\\
   & 0.65 & 36 & mt & 2.750(5) & 0.02061(11) & 85(3) \\

   & 0.67 & 36 & mt & 3.128(6) & 0.01986(12) & 150(5) \\
   & 0.67 & 36 & ov   & 3.127(2) & 0.01993(7) & 12.7(3) \\

   & 0.70 & 45 & mt & 3.787(6) & 0.01911(13) & 367(17) \\  
   & 0.70 & 30 & ov & 3.898(2)   & 0.01926(7) & 26.9(4)\\  
   & 0.70 & 40 & ov & 3.795(2)   & 0.01906(6) & 27.3(8)\\  
   & 0.70 & 45 & ov & 3.7885(10) & 0.01914(5) & 26.7(4)\\  
   & 0.70 & 50 & ov & 3.790(2)   & 0.01916(7) & 27.1(1.1)\\  

   & 0.72 & 54 & ov & 4.304(2)& 0.01875(6) & 44.3(1.6)\\

   & 0.75 & 60 & mt & 5.201(10)& 0.01851(20) & 1550(150)\\
   & 0.75 & 60 & ov & 5.195(3)   & 0.01815(8)  & 98(4)\\
   & 0.75 & 66 & ov & 5.198(4)   & 0.01821(16) & 99(3)\\
   & 0.75 & 72 & ov & 5.199(3)   & 0.01836(12) & 99(5)\\

   & 0.80 & 80 & ov & 7.091(3)   & 0.01757(12) & 420(20)\\ 
   & 0.80 & 90 & ov & 7.087(6)   & 0.01772(22) & 435(40)\\ 
   & 0.80 &100 & ov & 7.090(6)   & 0.01752(24) & 394(20) \\

   & 0.85 &120 & ov & 9.651(5)  & 0.01729(24) &1900(100)\\
   & 0.85 &140 & ov & 9.653(11) & 0.0180(5) & 1900(200)\\

   &0.87  & 150& ov & 10.904(10)& 0.0177(5) & 4800(700)\\
\hline

15 & 0.54 & 25 & ov & 1.7013(7) & 0.01382(4) & 6.4(2) \\

   & 0.56 & 30 & ov & 1.9409(9) & 0.01322(5) & 10.9(3) \\

   & 0.58 & 36 & ov & 2.2126(9) &  0.01266(5) & 19.5(4) \\

   & 0.60 & 42 & ov & 2.5185(12) & 0.01221(7) & 35.7(7) \\

   & 0.63 & 45 & ov & 3.050(3) & 0.01180(14) & 90(6) \\
   & 0.63 & 50 & ov & 3.058(3) & 0.01191(10) & 89(6) \\

   & 0.65 & 45 & ov & 3.462(3)  & 0.01128(11) & 193(5) \\
   & 0.65 & 54 & ov & 3.4630(11)& 0.01150(7) & 198(9) \\

   & 0.67 & 60 & ov & 3.9269(14)& 0.01135(5) & 400(20) \\

   & 0.70 & 75 & ov & 4.735(3) & 0.01137(16) & 1330(80) \\

   & 0.72 & 85 & ov & 5.361(3) & 0.01158(25) & 3100(300) \\ 

   & 0.73 & 90 & ov & 5.705(3)& 0.01132(20) & 5400(600) \\
\hline 

21  & 0.49 & 24 &  ov & 1.4203(5) & 0.00953(3) & 10.6(2) \\

    & 0.51 & 28 &  ov & 1.6208(4) & 0.00900(3) & 20.4(4) \\

    & 0.54 & 30 & ov & 1.9700(6) & 0.00845(4) & 61.0(1.4) \\
    & 0.54 & 36 & ov & 1.9715(6) & 0.00843(5) & 60.4(1.9) \\

    &0.57 & 30 & ov & 2.3887(8) & 0.00809(7) & 204(9) \\
    &0.57 & 38 & ov & 2.3893(9) & 0.00802(7) & 200(6) \\
    &0.57 & 42 & ov & 2.3895(9) & 0.00800(7) & 201(9)  \\

    & 0.60 & 45 & ov & 2.8850(9) & 0.00807(8)  & 820(30) \\
    & 0.60 & 48 & ov & 2.8884(6) & 0.00776(8)  & 770(30) \\
    & 0.60 & 50 & ov & 2.888(2)  & 0.00796(10) & 740(40) \\ 

    &0.62  & 54 & ov & 3.275(2) & 0.00777(15) & 2160(150) \\
    &0.62  & 56 & ov & 3.278(2) & 0.00789(10) & 1980(130) \\
    &0.62  & 60 & ov & 3.277(2) & 0.00807(22) & 1900(200) \\ 

    &0.64 & 60 & ov & 3.712(2) & 0.00800(14) & 5670(220) \\
    &0.64 & 64 & ov & 3.709(2) & 0.00798(16) & 6100(250) \\ 

    &0.66 & 72 & ov & 4.209(2) & 0.00807(25) & 19000(3000)\\
\hline\hline
\end{tabular}
\end{center}
\end{table*}

\begin{figure}[tb]
\centerline{\psfig{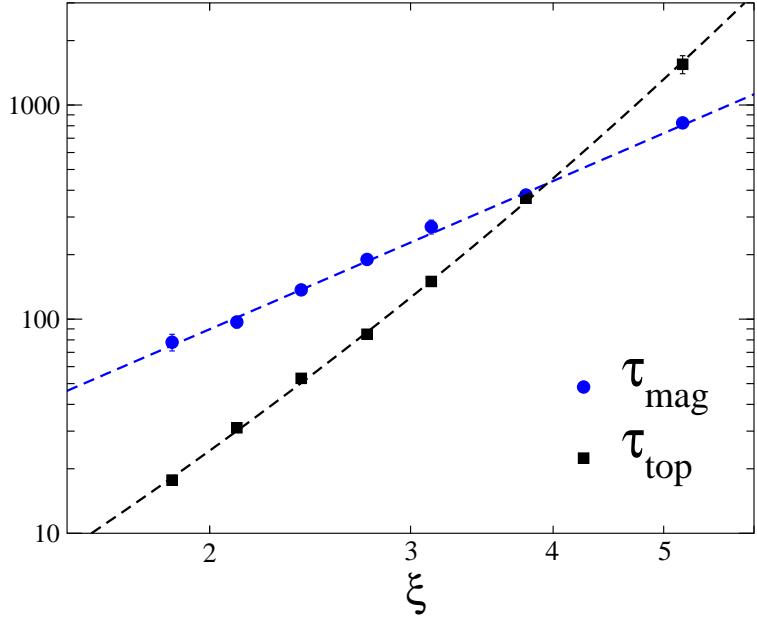}}
\vspace{2mm}
\caption{ Integrated autocorrelation time of the magnetic
susceptibility and the topological susceptibility for the $\mathrm{CP}^9$ model
and Metropolis updating.  The lines are the results of the fits
described in the text, power-law and exponential fits for $\tau_{\rm
mag}$ and $\tau_{\rm top}$ respectively.  }
\label{taumt}
\end{figure}

Let us first consider the results of the standard Metropolis algorithm
(50\% acceptance, 10 hits per lattice variable).  Figure~\ref{taumt}
shows the results for the integrated autocorrelation times of the
magnetic and topological susceptibilities obtained for the $\mathrm{CP}^9$
model.  The autocorrelation time $\tau_{\rm mag}$ of $\chi_m$ is in
agreement with the expected power-law behavior, i.e. $\tau_{\rm
mag}=c\xi^z$ with $z$ slightly larger than 2 (a fit of all data to
$\tau_{\rm mag}= c\xi^z$ gives $z=2.30(5)$ with $\chi^2/{\rm
d.o.f.}\simeq 0.9$).  On the other hand, the autocorrelation time
$\tau_{\rm top}$ of $\chi_t$ appears to increase much faster.  In
particular, a power-law behavior with $z\approx 2$ can be definitely
excluded.  The data for the largest available $\xi$ suggest larger
values of $z$, i.e. $z\gtapprox 4$. Moreover, an exponential ansatz
$\tau_{\rm top}\sim \exp ( c\xi^\theta)$ turns out to be well fitted
by all data, with $\theta\approx 0.3$, as shown in Fig.~\ref{taumt}.

However, in order to obtain a more precise characterization of the
topological CSD, data for larger values of $\xi$ are necessary, and
this becomes rather expensive when using a standard Metropolis
algorithm. A substantial improvement is obtained by using a more
effective local updating algorithm, constructed by employing also
overrelation procedures.  At each site the upgrading method was chosen
stochastically between an overrelaxed microcanonical (80\%), an
over-heat bath \cite{PV-91} (16\%), and the Metropolis algorithm (4\%)
to ensure ergodicity. Some details on the application of the above
algorithms to lattice $\mathrm{CP}^{N-1}$ models can be found in
Ref.~\cite{CRV-92}.  Similar mixtures are usually employed to obtain
effective local updating algorithms for 4-$d$ SU($N$) gauge theories.
The above mixed algorithm turns out to be much more effective than the
standard Metropolis.  For example for $N=10$ and $\beta=0.70$
($\xi=3.8$), we found $\tau_{\rm mag}\approx 380$ and $\tau_{\rm
top}\approx 367$ using the Metropolis algorithm, and $\tau_{\rm
mag}\approx 7$ and $\tau_{\rm top}\approx 27$ using the above
overrelaxed updating.  In addition, the mixed algorithm requires less
computer time by approximately a factor 2.  This allowed us to
obtain reliable estimates of $\tau_{\rm top}$ up to $\xi\simeq 10$ for
$N=10$ with a reasonable amount of computer time.  Actually, we do not
exclude that a further improvement can be achieved by optimizing the
mixture, since we did not really perform a detailed study of this
issue.~\footnote{ Optimization of overrelaxed algorithms is discussed
in Ref.~\cite{Wolff-92}.  } Moreover, we performed simulations for
larger values of $N$, $N=15$ and $N=21$, which will be useful to
understand the behavior of $\tau_{\rm top}$, and to compare $\chi_t$
with the available large-$N$ results.  The results are reported in
Table~\ref{tableres}.

\begin{figure}[tb]
\centerline{\psfig{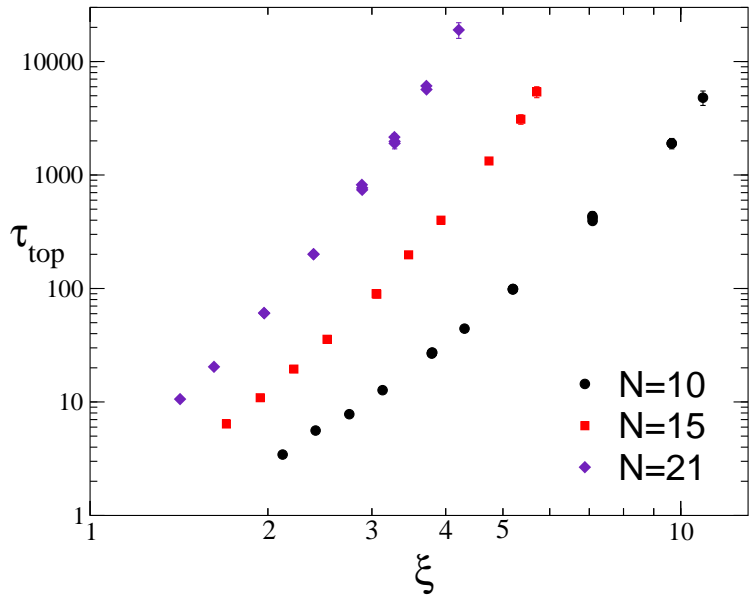}}
\vspace{2mm}
\caption{ Log-log plot of the integrated autocorrelation time
$\tau_{\rm top}$ versus $\xi$, for the $\mathrm{CP}^9$,
$\mathrm{CP}^{14}$ and $\mathrm{CP}^{20}$ models, obtained using the
mixed overrelaxed algorithm described in the text.  }
\label{tauov}
\end{figure}

Using the above random mixture of algorithms, the quasi-Gaussian modes
are expected to be still characterized by power-law CSD with $z\approx
2$. This is substantially confirmed by our simulations.  The data for
the autocorrelation time $\tau_{\rm top}$ of $\chi_t$ are shown in
Fig.~\ref{tauov}.  It is already apparent from the log-log plot of
$\tau_{\rm top}$ versus $\xi$ that the data do not agree with a simple
power law, i.e.  with $\tau_{\rm top}\approx c\xi^z$, on the whole
range of $\xi$ explored by this work. Moreover, even assuming an
asymptotic power-law behavior that sets at relatively large $\xi$,
values $z\approx 2$ can be definitely excluded, but substantially
larger $z$ are suggested by the data for the largest $\xi$.  In the
case of $N=10$ the behavior of $\tau_{\rm top}$ looks qualitatively
similar to the one found using the Metropolis algorithm.

\begin{figure}[tb]
\centerline{\psfig{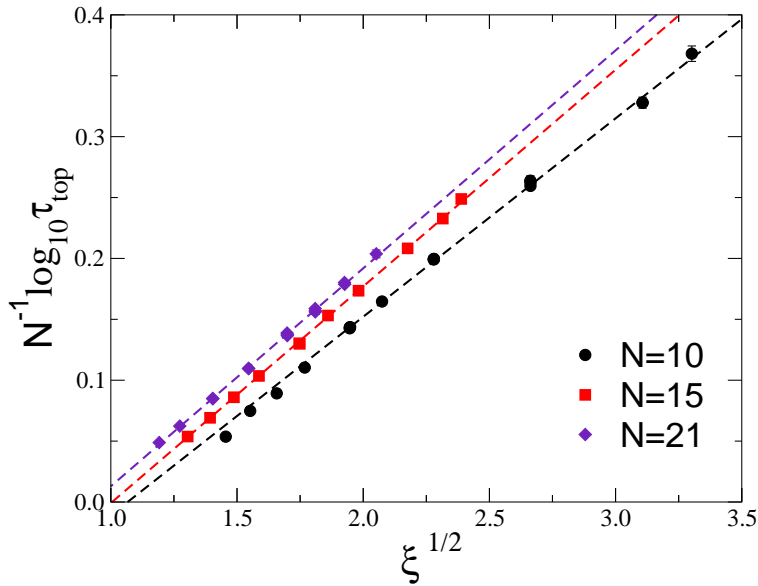}}
\vspace{2mm}
\caption{
The quantity $N^{-1}\log_{10}\tau_{\rm top}$ versus $\xi^{-1/2}$.
The lines show the results of the exponential fits $\tau_{\rm top}\sim \exp (b_N\xi^\theta)$
with $\theta=1/2$; see text.
}
\label{tauovsisq}
\end{figure}

Comparing the data of $\tau_{\rm top}$ at different $N$ but fixed
$\xi$, we note that the quantity $N^{-1}\log_{10}\tau_{\rm top}$ seems
to converge to a non-trivial large-$N$ limit, the approach being
roughly $O(1/N)$, see also Fig.~\ref{tauovsisq}.  This fact is also
suggested by the following simple picture.  Let us assume that the
transition from one topological sector to the other happens by
tunnelling through a potential barrier.  The resulting autocorrelation
time is $\tau\sim \exp S_b$ (neglecting entropy), where $S_b$ is the
action of the typical configurations that are at the boundary of the
different topological sectors.  Let us also assume that these
configurations are instanton-like.  Since the instanton action is
given by $S_I=N 2\pi/g(\rho)$ \cite{instanton} where $\rho$ is the
size of the instanton and $g(\rho)$ the running coupling at scale
$\rho$, we should expect that $\ln \tau_{\rm top} = O(N)$.  Note that
the same arguments apply to 4-$d$ SU($N$) lattice gauge theories, and
the estimates of the topological autocorrelation times for $N=3,4,6$
\cite{DPV-02,LT-01} are indeed consistent with the above dependence.

Proceeding further within this instanton picture, one arrives at a
power-law behavior with $z\sim N$.  Let us further assume that the
size of the relevant instanton configurations at the boundary of
different topological sectors is given by $\rho\approx a$ (see, e.g.,
Refs.~\cite{LT-01,BBHN-96}).  Then, we should have $g(a)\sim 1/\beta$
and using asymptotic freedom $\xi \sim \exp(2\pi\beta)$, thus
$\tau_{\rm top}\sim \exp S_I \sim \xi^z$ with $z\sim N$.  As is
already apparent from Fig.~\ref{tauov}, a reasonable fit to a simple
power law
\begin{equation}
\tau_{\rm top} = b_N \xi^{c_N N}
\label{powans}
\end{equation}
can be obtained only by discarding several data points at the smallest
values of $\xi$.  As Fig.~\ref{tauov} already shows, the fitted value
of $c_N$ tends to increase when more and more data at small $\xi$ are
discarded. We can tentatively determine lowest bounds for the power
coefficient $c_N$ by using only the largest values of $\xi$.  For
$N=10$, using the data for the last three $\beta$-values, i.e. data
for $\xi\gtapprox 7$, we obtain $c_{10}=0.51(2)$, $b_{10}=0.02(1)$.
In this case, we also note that there is a hint of stability in the
results for $c_{10}$, because a consistent value is already obtained
taking data for $\xi>5$, with an acceptable $\chi^2$.  In the case
$N=15$, the last three $\beta$-values give $c_{15}=0.49(4)$,
$b_{15}=0.02(2)$.  Finally, for $N=21$ the data for the last three
$\beta$-values give $c_{21}=0.41(2)$, $b_{21}=0.07(4)$.  We note that
these results for $c_N$ are rather close, consistently with the
expectation that $c_N=O(1)$ in the large-$N$ limit.  Consistent
results are obtained by considering a more general ansatz such as
$\tau_{\rm top} = a_N + b_N \xi^{c_N N}$, where we also allow for a
constant term.  Note that the naive guess obtained simply by assuming
$g(\rho=a)=1/\beta$ would be $z=N$. In conclusion, assuming a
power-law CSD, this analysis indicates that $z\gtapprox N/2$.

On the other hand, Fig.~\ref{tauovsisq} is also suggestive of an
exponential behavior, which emerges naturally from tunnelling through
a free-energy barrier whose size scales like $\xi^\theta$. Such
exponential behavior gives a good description of the data in the whole
range of $\xi$ explored by this work.  Therefore, we also consider an
exponential ansatz
\begin{equation}
{1\over N} \log_{10} \tau_{\rm top} = a_N + b_N \xi^{\theta_N},
\label{expans}
\end{equation}
where $a_N$, $b_N$ and $\theta_N$ are $O(1)$ in the large-$N$ limit,
and $\theta_N\approx 1/2$.  We may further simplify this ansatz by
assuming that $\theta_N$ is independent of $N$,
i.e. $\theta_N=\theta$. A global fit to the data gives
$\theta=0.49(2)$.  This result was obtained by discarding a few data
points at the smallest $\xi$ for each $N$, i.e. taking the data for
$\xi\gtapprox 3$ ($\beta \ge 0.70$) at $N=10$, $\xi\gtapprox 2$ at
$N=15$ and $N=21$ (corresponding to $\beta\ge 0.56$ and $\beta\ge
0.54$ respectively), in order to obtain an acceptable $\chi^2/{\rm
d.o.f.}\approx 0.9$.  In Fig.~\ref{tauovsisq} we also show the curve
obtained by fits with $\theta=1/2$, for which, using the same data as
in the above global fit, we obtained $a_{10} = -0.174(3)$ and
$b_{10}=0.163(2)$ with $\chi^2/{\rm d.o.f.}\approx 1.1$, $a_{15} =
-0.179(3)$ and $b_{15}=0.178(2)$ with $\chi^2/{\rm d.o.f.}\approx
0.4$, $a_{21} = -0.169(2)$ and $b_{21}=0.181(2)$ with $\chi^2/{\rm
d.o.f.}\approx 1.0$.

In conclusion, the CSD of the topological modes in Monte Carlo
simulation employing local updating algorithms turns out to be much
stronger than the one experienced by quasi-Gaussian modes. This has
been inferred by comparing the integrated autocorrelation time of the
magnetic and topological susceptibilities as a function of $\xi$.
Their behavior suggests an effective separation of short-time
relaxation within the topological sectors from long-time relaxation
related to the transitions between different topological sectors.  A
heuristic explanation can be devised by assuming the presence of
significant free-energy barriers in the configuration space between
different topological sectors, with the system changing topology by
tunnelling through such barriers. An exponential ansatz,
i.e. $\tau_{\rm top}\sim \exp (c\xi^\theta)$ with $\theta\approx 1/2$,
provides a good effective description of the data in the range of
$\xi$ where data are available.  However, the statistical analysis of
the available data for $\tau_{\rm top}$ does not actually allow us to
distinguish between an exponential CSD and an asymptotic power-law
behavior with $z\gtapprox N/2$ setting at relatively large
$\xi$. Power-law behaviors with smaller exponents can definitely be
excluded by our analysis. Data for larger $\xi$ and/or a better
modellization of the Monte Carlo dynamics of the topological modes
would be needed to further clarify this issue. We argue that the
severe CSD experienced by the topological modes under local updating
algorithms should be a general feature of Monte Carlo simulations of
lattice models with non-trivial topological properties, since the
mechanism behind this phenomenon should be similar.  This is also
supported by recent Monte Carlo simulations of 4-$d$ lattice SU($N$)
gauge theories reported in Refs.~\cite{DPV-02,LT-01}.  Indeed, the
estimates of autocorrelation time $\tau_Q$ of the topological
charge~\footnote{
In our simulation of $\mathrm{CP}^{N-1}$ models we also measured the
autocorrelation time $\tau_Q$ of the topological charge
(\ref{qgeomP}). In all cases we found $\tau_Q/\tau_{\rm top} \simeq 2$
(more precisely $\tau_Q/\tau_{\rm top} \simeq 2.3$ and
$\tau_Q/\tau_{\rm top} \simeq 2.1$ respectively for the Metropolis and
overrelaxed simulations).  Note that a simple Gaussian propagation
would give $\tau_Q/\tau_{\rm top}=2$.}
recently reported in Ref.~\cite{DPV-02} (measured using the cooling
technique) showed a rapid increase with the length scale, and an
apparent exponential behavior $\tau_Q\sim \exp (c\xi)$ in the range of
values of $\xi$ where data were available, for all $N=3,4,6$.  We
stress again that the CSD of topological modes may represent a serious
limitation for simulations of lattice QCD, in order to study physical
issues determined by the topological excitations, such as the physics
of the $\eta'$ meson. Our results suggest that the contribution of the
correlation time to the total cost of a simulation could be higher
than is usually assumed, if one wants to sample the different
topological sectors correctly. In particular, it may worsen the
current cost estimates of the dynamical fermion simulations for
lattice QCD, see e.g. Ref.~\cite{Jansen-03}, where it is usually
assumed that the autocorrelation time only contributes a factor of
$\xi$.

An interesting question would be whether other, possibly non-local,
updating algorithms may eliminate or at least improve the severe form
of CSD of topological modes.  Cluster algorithms turn out not to be
effective in $\mathrm{CP}^{N-1}$ models \cite{JW-92,CEPS-93}.
Instead, as shown in Ref.~\cite{HM-92}, multigrid Monte Carlo
algorithms achieve a substantial reduction of the CSD of the
quasi-Gaussian modes relevant to the magnetic susceptibility. It is
however not clear if they can also accelerate the decorrelation of the
topological modes. Let us mention here that algorithms based on the
simulated tempering method \cite{MP-92} were also tried in
Ref.~\cite{V-93}, but apparently without achieving a particular
advantage.

\begin{figure}[tb]
\centerline{\psfig{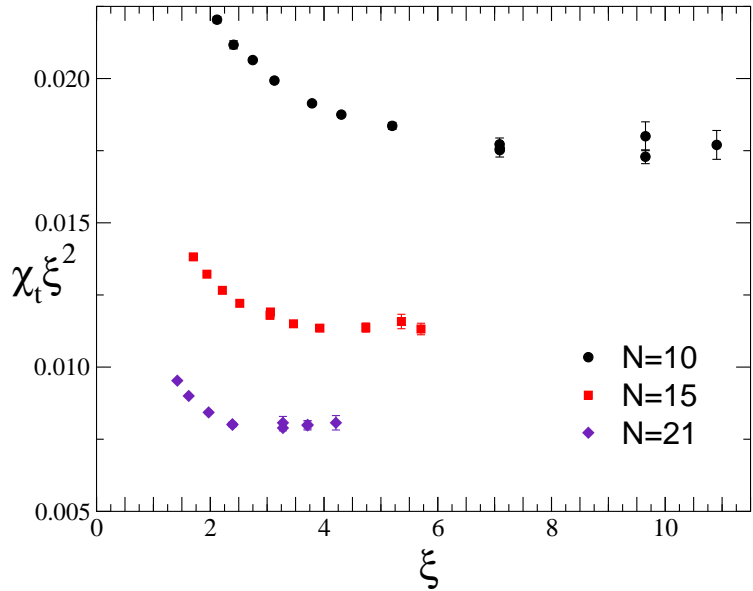}}
\vspace{2mm}
\caption{
$\chi_t\xi^2$ versus $\xi$ for $N=10,15,21$.
}
\label{chitfig}
\end{figure}

Finally let us discuss the results for the topological susceptibility;
data for the dimensionless quantity $\chi_t\xi^2$, reported in
Fig.~\ref{chitfig}, clearly show a plateau for the largest values of
$\xi$, where one can extract an estimate of the continuum limit of
$\chi_t\xi^2$.  We obtain:
\begin{eqnarray}
&& \chi_t \xi^2 = 0.0175(3) \quad {\rm for}\quad N=10,
\label{chitest} \\
&& \chi_t \xi^2 = 0.0113(2) \quad {\rm for}\quad N=15,
\nonumber \\
&& \chi_t \xi^2 = 0.0080(2) \quad {\rm for}\quad N=21,
\nonumber
\end{eqnarray}
where, prudently, we have taken the typical error on the data at the
plateau as estimate of the uncertainty.  Compatible but substantially
less precise results for $N=10,21$ are reported in
Refs.~\cite{CRV-92,V-93}.  The estimates (\ref{chitest}) may be
compared with the available results obtained in the framework of the
large-$N$ expansion \cite{CR-91}:
\begin{equation}
\chi_t\xi^2 = {1\over 2\pi N} - {0.060\over N^2} + O(1/N^3).
\end{equation}
We note that the estimates (\ref{chitest}) are slightly larger.
Actually they suggest a $O(1/N^2)$ contribution given by $c_2/N^2$
with $c_2\simeq 0.2$.  Indeed, by evaluating the quantity
$N^2(\chi_t\xi^2 - \case{1}{2\pi N})$, using the estimates
(\ref{chitest}), one would obtain $c_2=0.16(3)$ for $N=10$,
$c_2=0.16(4)$ for $N=15$, and $c_2=0.19(9)$ for $N=21$.  However, this
apparent discrepancy can be easily accounted for by a slow approach to
the large-$N$ regime, and the apparent stability of the $O(1/N^2)$
correction may be only a chance.  This point deserves further
investigation.

\bigskip

{\bf Acknowledgements} We thank Maurizio Davini for his valuable and
indispensable technical support.  We also thank Philippe de Forcrand,
Martin L\"uscher, Andrea Pelissetto, and Paolo Rossi for interesting
and useful discussions.

\end{document}